# A Guide to Design Disturbance Observer based Motion Control Systems

Emre SARIYILDIZ, Kouhei OHNISHI
Department of System Design Engineering
Keio University, Yokohama, Japan
emre@sum.sd.keio.ac.jp, ohnishi@sd.keio.ac.jp

***Abstract*- This paper proposes new practical design tools for the robust motion control systems based on disturbance observer (DOB). Although DOB has long been used in several motion control applications, it has insufficient analysis and design tools. The paper proposes new practical robustness constraints, which improve the robustness at high frequencies, on the bandwidth of a DOB and nominal inertia. Although increasing the bandwidth of a DOB and nominal inertia improves the performance and stability, they are limited by the robustness constraint. Besides, a novel stability analysis method is proposed for reaction force observer (RFOB) based robust force control systems. It is shown that not only the performance, but also the stability changes significantly by the imperfect identification of inertia and torque coefficient. The robustness and stability of a DOB based motion control system are improved by proposing new design tools. The validity of the proposals are verified by experimental results.**

***Index Terms: Disturbance Observer, Motion Control Systems, Reaction Force Observer, Robustness and Stability***

## I. INTRODUCTION

Three decades before, DOB was proposed by Ohnishi et al. in the first IPEC conference [1]. After that, it has been used in several robust motion control applications such as robotics, industrial automation, automotive, and so on [2-4]. A DOB estimates external disturbances and system uncertainties, e.g., friction, inertia variation etc.; and the robustness of the motion control system is achieved by feeding back the estimated disturbances in an inner-loop [5]. To achieve control goals, e.g., position, force or admittance control goals, an outer-loop controller is designed by considering the nominal plant parameters, since a DOB can nominalize an uncertain plant [5].

A low-pass-filter (LPF), which satisfies the properness in the inner-loop, and the nominal inertia and torque coefficient of a motor are used in the design of a DOB. The dynamic characteristics of the DOB's LPF and the ratio between the uncertain and nominal plant parameters change the stability and robustness of a DOB based motion control system significantly [6]. It is a well-known fact that the bandwidth of the DOB's LPF is desired to set as high as possible to estimate/compensate disturbances in a wide frequency range. However, its bandwidth is limited by practical, i.e., noise, and robustness constraints [7]. Besides that the order of the DOB's LPF changes the robustness and performance significantly. As it is increased, the performance of the system improves, yet the robustness deteriorates [7]. The stability and performance of a DOB based robust motion control system can be improved by using higher/lower nominal inertia/torque coefficient; however, its upper/lower bound has not been derived yet [8].

A RFOB, which was proposed by Murakami et al., is an application of a DOB and is used to estimate external loading forces [9]. It has several superiorities over a force sensor such as sensorless force control, higher force control bandwidth, etc.; and they have been shown in the literature experimentally [10, 11]. A RFOB is designed easily by subtracting the system uncertainties from the input of a DOB. Therefore, a DOB and a RFOB are quite similar, however only the latter has a model based control structure. It is the main challenging issue in a RFOB based robust force control system, since imperfect system identification may change the stability and performance significantly. So far, only the performance of a RFOB based robust force control system has been considered in the literature due to the oversimplified analysis methods. However, not only the performance, but also the stability changes significantly by the design parameters of a DOB and a RFOB.

In this paper, new design constraints are proposed for DOB based robust position and force control systems. It is shown that the bandwidth of a DOB and nominal inertia have robustness constraints due to imperfect velocity measurement in practice. The proposed robustness constraint is more dominant than noise, so the bandwidth of a DOB should be determined by considering the proposed robustness constraint. The stability of a DOB based motion control system is improved by increasing/decreasing nominal inertia/torque coefficient, yet the robustness deteriorates. Therefore, there is a trade-off, which is adjusted by the ratio of the uncertain and nominal plant parameters, between the stability and robustness of a DOB based robust motion control system. A novel stability analysis method is proposed for RFOB based robust force control systems. It is shown that the stability of a RFOB based robust force control system deteriorates as the identified inertia that is used in the design of a RFOB is increased. Besides, it is shown that a DOB and a RFOB can be designed as a phase lead-lag compensator in the robust force control systems by setting the bandwidths to the different values.

The rest of the paper is organized as follows. In section II, a DOB and a RFOB are presented briefly. In sections III and IV, new practical design constraints are proposed for DOB and RFOB based position and force control systems, respectively. In section V, experimental results are given. The paper ends with conclusion given in the last section.

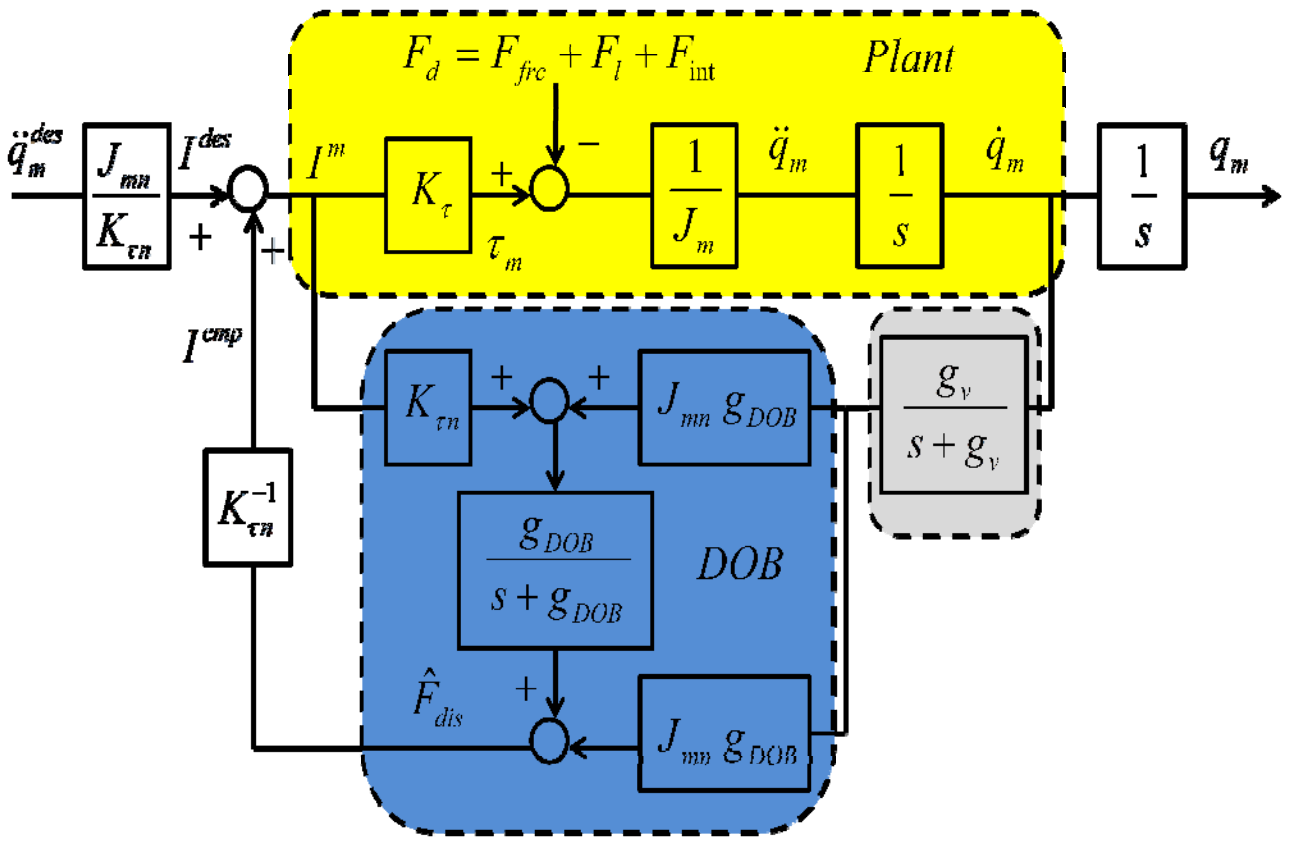

(a) Disturbance Observer

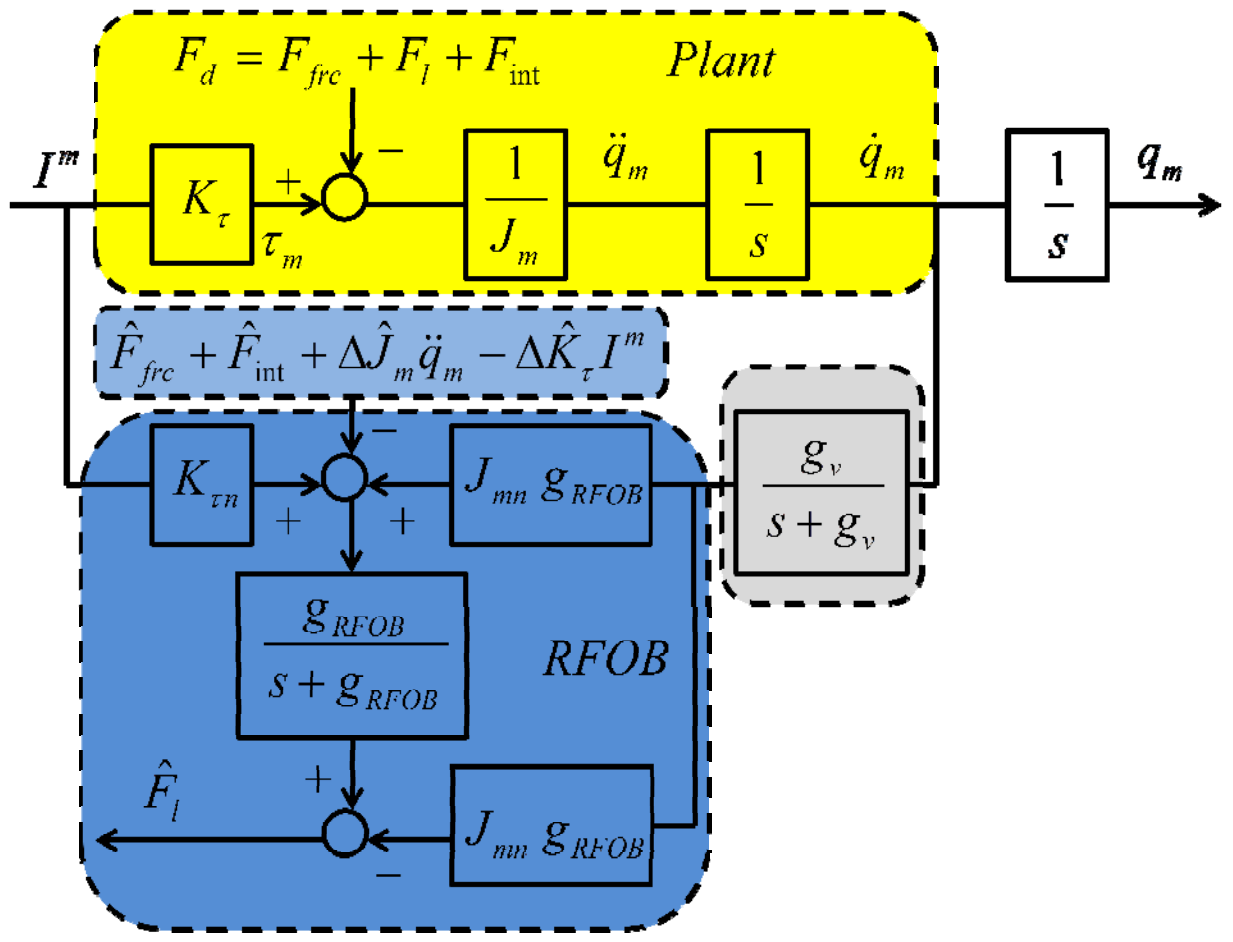

(b) Reaction Force Observer

Fig. 1 Block diagrams of a DOB and a RFOB

## II. DISTURBANCE AND REACTION FORCE OBSERVERS

Block diagrams of a DOB and a RFOB are shown in Fig. 1. In this figure

| | |
|---|---|
| $J_m, J_{mn}$ | Uncertain and nominal inertias; |
| $K_\tau, K_{\tau n}$ | Uncertain and nominal torque coefficients; |
| $I^m, I^{des}, I^{cmp}$ | Total, desired and compensate motor currents; |
| $q_m, \dot{q}_m, \ddot{q}_m$ | Angle, velocity and acceleration; |
| $g_{DOB}$ | Cut-off frequency of a DOB; |
| $g_{RFOB}$ | Cut-off frequency of a RFOB; |
| $g_v$ | Cut-off frequency of velocity measurement; |
| $\Delta J_m$ and $\Delta \hat{J}_m$ | Inertia variation and its estimation; |
| $\Delta K_\tau$ and $\Delta \hat{K}_\tau$ | Torque coefficient variation and its estimation; |
| $F_l$ | Loading force; |

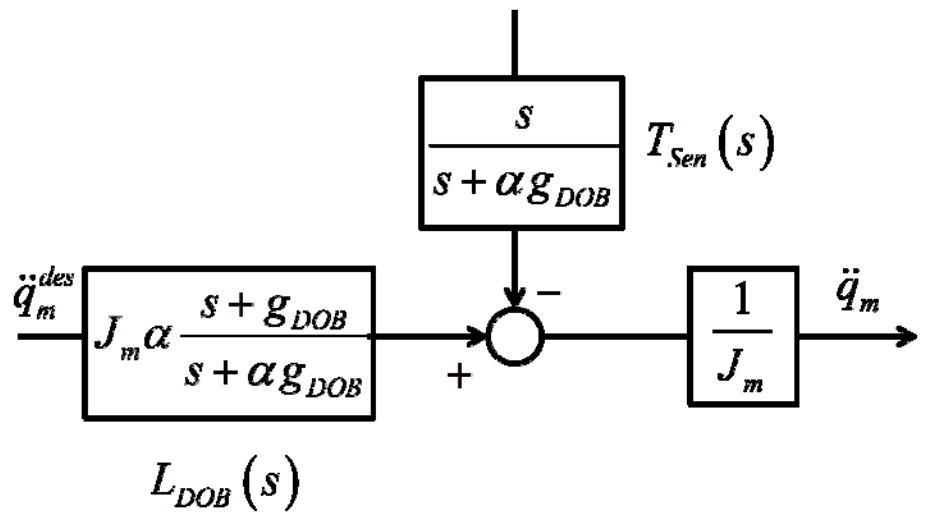

(a) Ideal velocity measurement

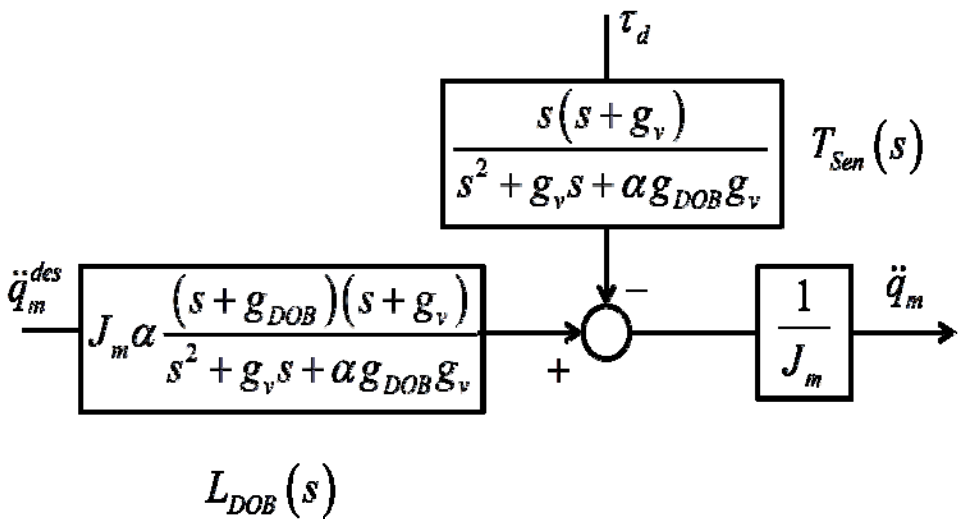

(b) Imperfect velocity measurement

Fig. 2 Simplified block diagrams of a DOB

| | |
|---|---|
| $F_{frc}$ | Nonlinear friction force; |
| $F_{int}$ | Interactive force; |
| $F_d = F_{int} + F_l + F_{frc}$ | Total external disturbance; |
| $F_{dis}, \hat{F}_{dis}$ | Total disturbance and its estimation; |

As shown in the Fig. 1, although a DOB and a RFOB are quite similar, only the latter has a model based control structure. Since, not only the nominal, but also the uncertain plant parameters are required in the design of a RFOB.

Without any approximation, the Fig. 1 (a) can be simplified as shown in the Fig. 2, in which $\alpha = \frac{J_{mn} K_\tau}{J_m K_{\tau n}}$. The Fig. 2 (a) is drawn when ideal velocity measurement is achieved, i.e., $g_v$ is infinite; however, the Fig. 2 (b) is drawn when $g_v$ is finite. It is a well-known fact that a DOB requires precise velocity measurement [12]. Therefore, in practice, $g_v$ should be finite to suppress noise and obtain precise velocity measurement in a determined bandwidth.

The simplified block diagrams given in the Fig.2 show the open loop $\left(L_{DOB}(s)\right)$ and sensitivity $\left(T_{Sen}(s)\right)$ transfer functions of a DOB based motion control system. The open loop transfer functions show that a DOB can be designed as a phase lead-lag compensator that is adjusted by $\alpha$. The stability and performance can be improved by using a DOB as a phase lead compensator, i.e., $\alpha > 1$. The dynamic characteristics of the sensitivity function changes significantly at high frequencies when $g_v$ is finite. Therefore, although it has never been considered so far, the robustness of a DOB changes significantly when a low-pass-filter is used in velocity measurement. A new robustness design constraint for a DOB based motion control system is derived as follows:

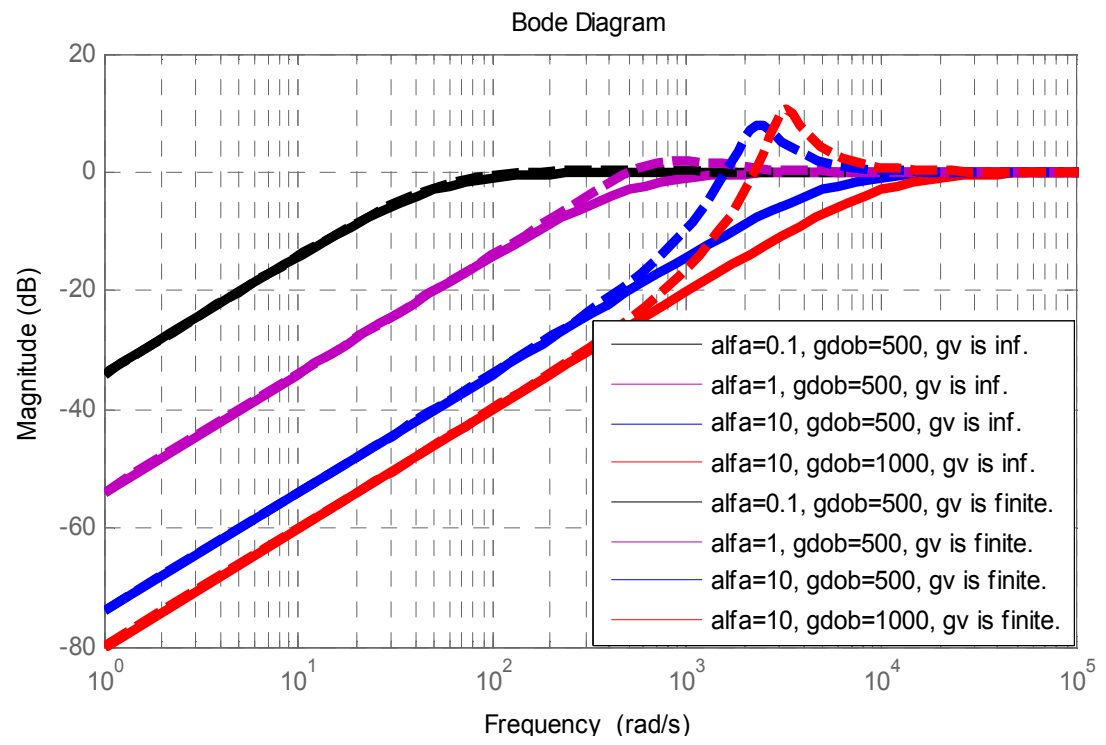

Fig. 3. Sensitivity function $(T_{Sen})$ frequency responses of an inner-loop

Let us consider the sensitivity function that is given in the Fig. 2 (b) and apply $g_v = \kappa g_{DOB}$.

$$T_{Sen} = \frac{s(s+\kappa g_{DOB})}{s^2 + \kappa g_{DOB} s + \alpha\kappa g_{DOB}^2} \tag{1}$$

The characteristic function of (1) can be designed by using

$$w_n = \sqrt{\alpha\kappa}\, g_{DOB} \text{ and } \xi = 0.5\sqrt{\frac{\kappa}{\alpha}} \tag{2}$$

where $w_n$ and $\xi$ denote natural frequency and damping coefficient of a general second order characteristic polynomial, respectively. To suppress the peak of the frequency response of $T_{Sen}$ , if it is assumed that $\xi \geq 0.707$, then

$$\alpha g_{DOB} \leq \frac{g_v}{2} \tag{3}$$

The equation (3) indicates that $\alpha$ and/or $g_{DOB}$ cannot be increased freely due to the robustness constraint. As a result, there is a trade-off between the stability and robustness in the design of a DOB. Fig.3 indicates the robustness constraint and the trade-off between the stability and robustness of a DOB based motion control system. Although the robustness of a DOB is improved as $\alpha$ is increased when $g_v$ is infinite, the peak of the sensitivity function increases, which deteriorates the robustness, as $\alpha$ is increased when $g_v$ is finite.

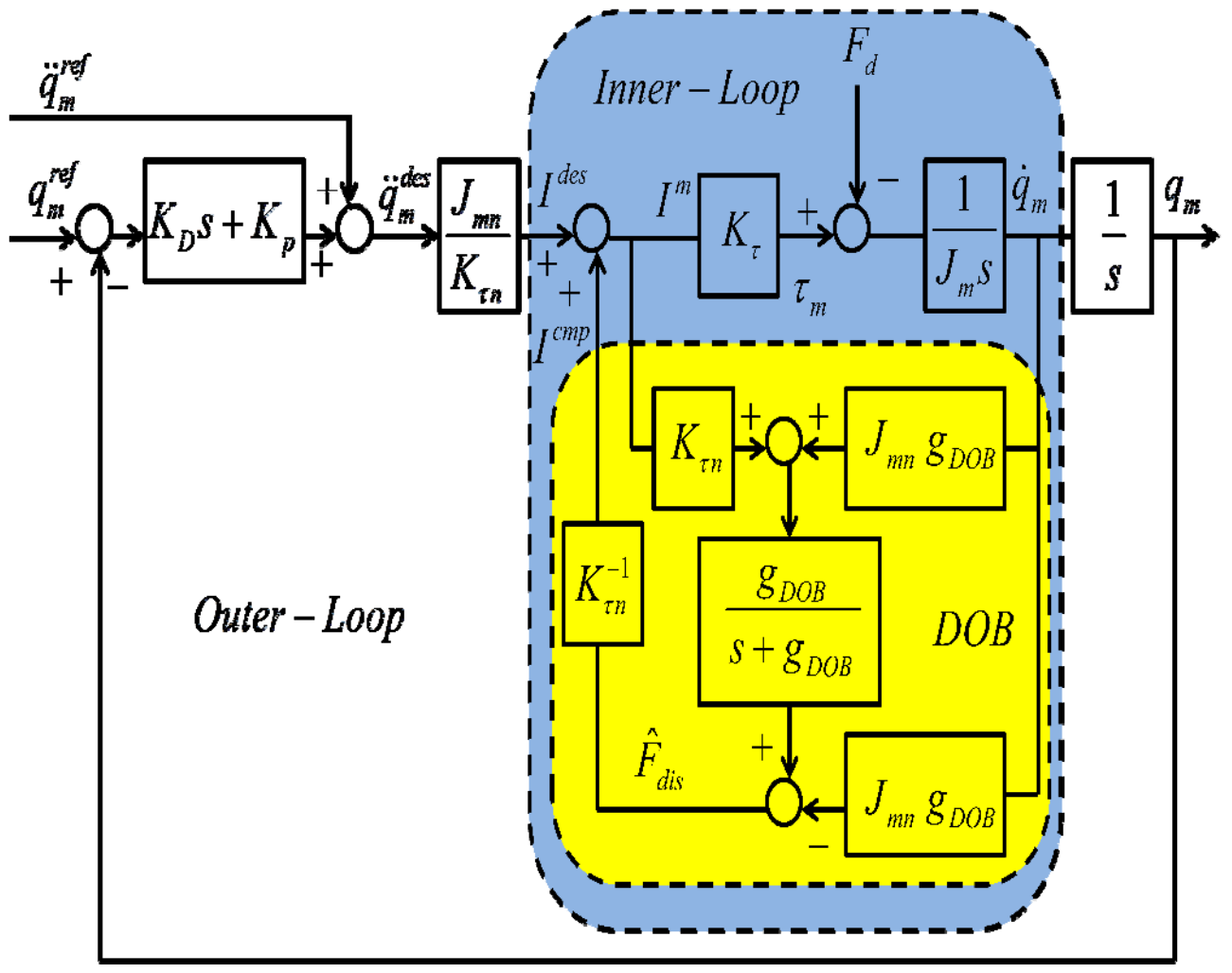

Fig. 4. A Block diagram for a DOB based robust position control system

## III. DOB Based Robust Position Control Systems

Fig. 4 shows a block diagram for a DOB based robust position control system. In this figure, $q_m^{ref}$ , $\dot{q}_m^{ref}$ and $\ddot{q}_m^{ref}$ denote angle, velocity and acceleration reference inputs, respectively; $\ddot{q}_m^{des}$ denotes the desired acceleration; and $K_P$ and $K_D$ denote the proportional and derivative gains of the outer-loop controller, respectively. A DOB provides the robustness of the position control system in the inner-loop, and the performance goals are achieved by using an acceleration based controller in the outer-loop. The transfer functions between $\ddot{q}_m^{ref}$ and $\ddot{q}_m$ can be derived from the Fig. 4 directly as follows:

$$\frac{\ddot{q}_m}{\ddot{q}_m^{ref}} = \frac{s(s+g_{DOB})(s^2+K_D s+K_p)}{\alpha^{-1}s^3 + (g_{DOB}+K_D)s^2 + (K_P + g_{DOB}K_D)s + g_{DOB}K_P} \tag{4}$$

when $g_v$ is infinite; and

$$\frac{\ddot{q}_m}{\ddot{q}_m^{ref}} = \frac{\alpha(s+g_v)(s+g_{DOB})(s^2+K_D s+K_p)}{s^4 + g_v s^3 + \alpha g_v g_{DOB} + \alpha(s+g_v)(s+g_{DOB})(s^2+K_D s+K_p)} \tag{5}$$

when $g_v$ is finite.

If the stabilities of the transfer functions given in (4) or (5) are analyzed, for instance the Routh-Hurwitz theorem can be used, then it can be shown that the stability of the position control system is improved by increasing $\alpha$ , i.e., using higher nominal inertia or lower nominal torque coefficient improve the stability of a DOB based position control system. However, the equation (3) shows that $\alpha$ cannot be increased freely due to the robustness constraint in practice.

## IV. RFOB Based Robust Force Control Systems

Fig. 5 shows a block diagram for a RFOB based robust force control system. In this figure, $C_f$ denotes the force control gain; $F_l^{ref}$ denotes the force reference; and $F_l$ and $\hat{F}_l$ denote external force and its estimation, respectively. The other parameters are same as defined above. The stability of a RFOB based robust force control system can be analyzed by deriving the open-loop transfer function as follows:

Environmental contact model can be described effectively by using a simple lumped spring damper model as follows:

$$F_l = D_{env}(\dot{q}_m - \dot{q}_e) + K_{env}(q_m - q_e) \tag{6}$$

where $D_{env}$ and $K_{env}$ denote environmental damping and stiffness coefficients, respectively; and $q_e$ and $\dot{q}_e$ denote the position and velocity of environment at equilibrium, respectively. The open

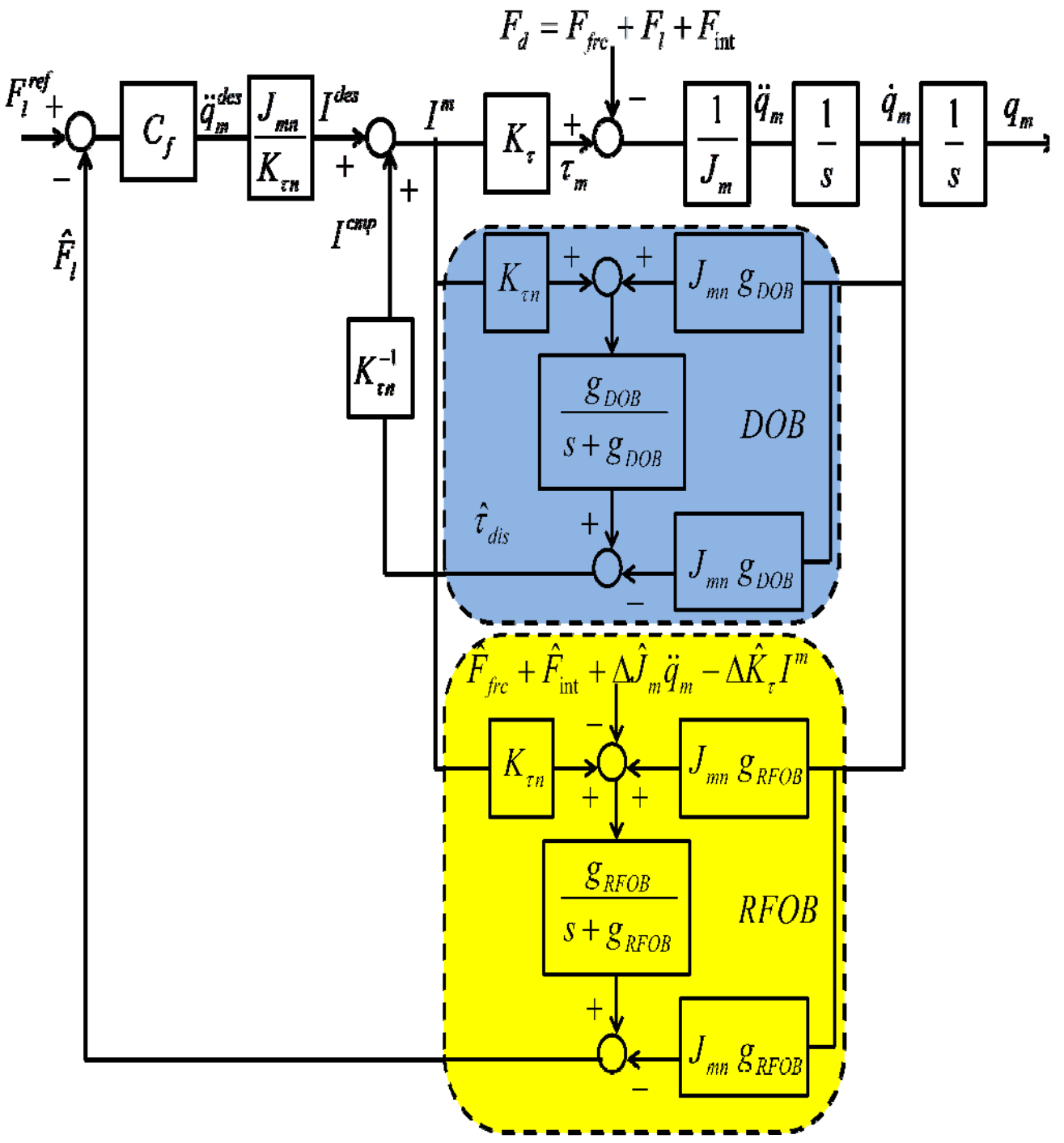

Fig. 5. A Block diagram for a RFOB based robust force control system

loop transfer function of a RFOB based robust force control system is derived from the Fig. 5 as follows:

$$L_{RFOB}(s) = C_f \frac{g_{RFOB}\dfrac{J_{mn}}{K_{\tau n}}(s+g_{DOB})\varphi(s)}{s\{J_m s(s+\alpha g_{DOB})+(D_{env}s+K_{env})\}(s+g_{RFOB})} \quad (7)$$

where $\varphi(s) = (J_m\hat{K}_\tau - \hat{J}_m K_\tau)s^2 + \hat{K}_\tau D_{env}s + \hat{K}_\tau K_{env}$. The equation (7) shows that if $J_m\hat{K}_\tau < \hat{J}_m K_\tau$, then the open loop transfer function has a zero at right half plane. Therefore, the stability and performance of a RFOB based robust force control system may deteriorate by the imperfect identification of inertia and torque coefficient, significantly. Consequently, a RFOB should be designed by satisfying $J_m\hat{K}_\tau > \hat{J}_m K_\tau$ to improve the stability of the system. Besides, the equation (7) shows that the relative degree of $L_{RFOB}(s)$ is one, so the root loci have asymptotes, at $\pi$.

If a RFOB is designed by using perfect system identification, i.e., $J_m\hat{K}_\tau = \hat{J}_m K_\tau$ , then

$$L_{RFOB}(s) = C_f \frac{(s+g_{DOB})}{(s+g_{RFOB})} \frac{g_{RFOB}J_m\alpha(D_{env}s+K_{env})}{s\{J_m s(s+\alpha g_{DOB})+(D_{env}s+K_{env})\}} \quad (8)$$

The equation (8) shows that a DOB and a RFOB can be designed as a phase lead-lag compensator, and the stability of the robust force control system can be improved by using $g_{DOB} < g_{RFOB}$. When perfect system identification is achieved, the relative degree of $L_{RFOB}(s)$ is two, so the root loci have asymptotes, at $\pm\pi/2$. The equations (7) and (8) show that the asymptotic behaves of the root loci deteriorate by the perfect inertia and torque coefficient identification.

In general, the bandwidths of a DOB and a RFOB are set to the same value in the robust force control systems. If $g_{DOB} = g_{RFOB} = g$, then the open-loop transfer function is

$$L_{RFOB}(s) = C_f \frac{gJ_m\alpha(D_{env}s+K_{env})}{s\{J_m s(s+\alpha g)+(D_{env}s+K_{env})\}} \quad (9)$$

The relative degree of $L_{RFOB}(s)$ is two, so the root loci have asymptotes, at $\pm\pi/2$ .However, the phase lead-lag compensator cannot be used in the design of the robust force control systems.

The equations (7), (8) and (9) show that each of the open loop transfer functions have a pole at the origin, so there is no a steady state error in the DOB based robust force control systems.

Table I
Specifications of the experimental setup

| | |
|---|---|
| $m \cong 0.62 kg$ | Nominal mass |
| $K_{\tau n} = 33 N/A$ | Nominal force coefficient |
| $g_v = 1000 rad/s.$ | Cut-off frequency of vel. meas. |
| $g_{DOB} = 250 rad/s.$ | Cut-off frequency of DOB. |
| $g_{RTOB} = 750 rad/s.$ | Cut-off frequency of RFOB |
| $K_p = 1200$ | Proportional position control gain |
| $K_D = 90$ | Derivative position control gain |
| $C_f = 1$ | Proportional force control gain |

## V. Simulation and Experiment

In this section, simulation and experimental results will be presented. In the experiments, a linear DC motor which is shown in Fig. 6 is used. Specifications of the experimental set-up are shown in Table-I. The sampling time is 0.1 ms, and KYOWA LUR-A-50NSA1 force sensor is used to verify the performance of RFOB in force control.

Fig. 7 and Fig. 8 show the root loci of a DOB based position and force control systems, respectively. The root locus of the position control system is plotted with respect to $\alpha$ in the Fig.

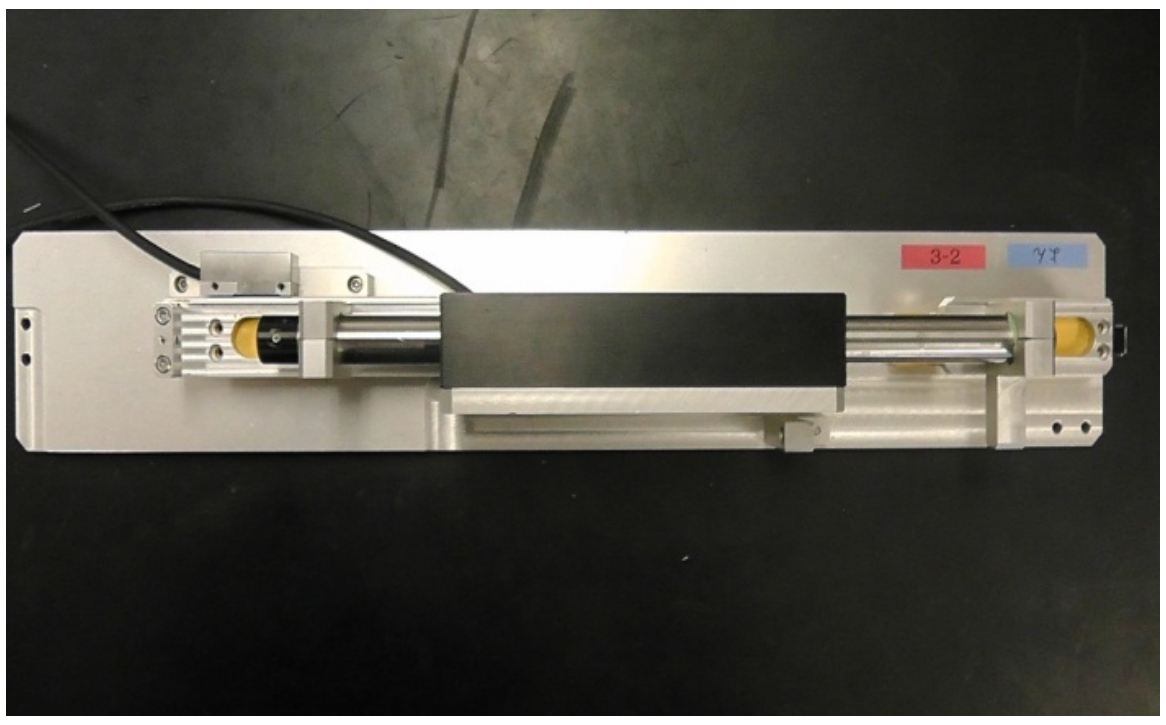

Fig. 6. Linear DC motor

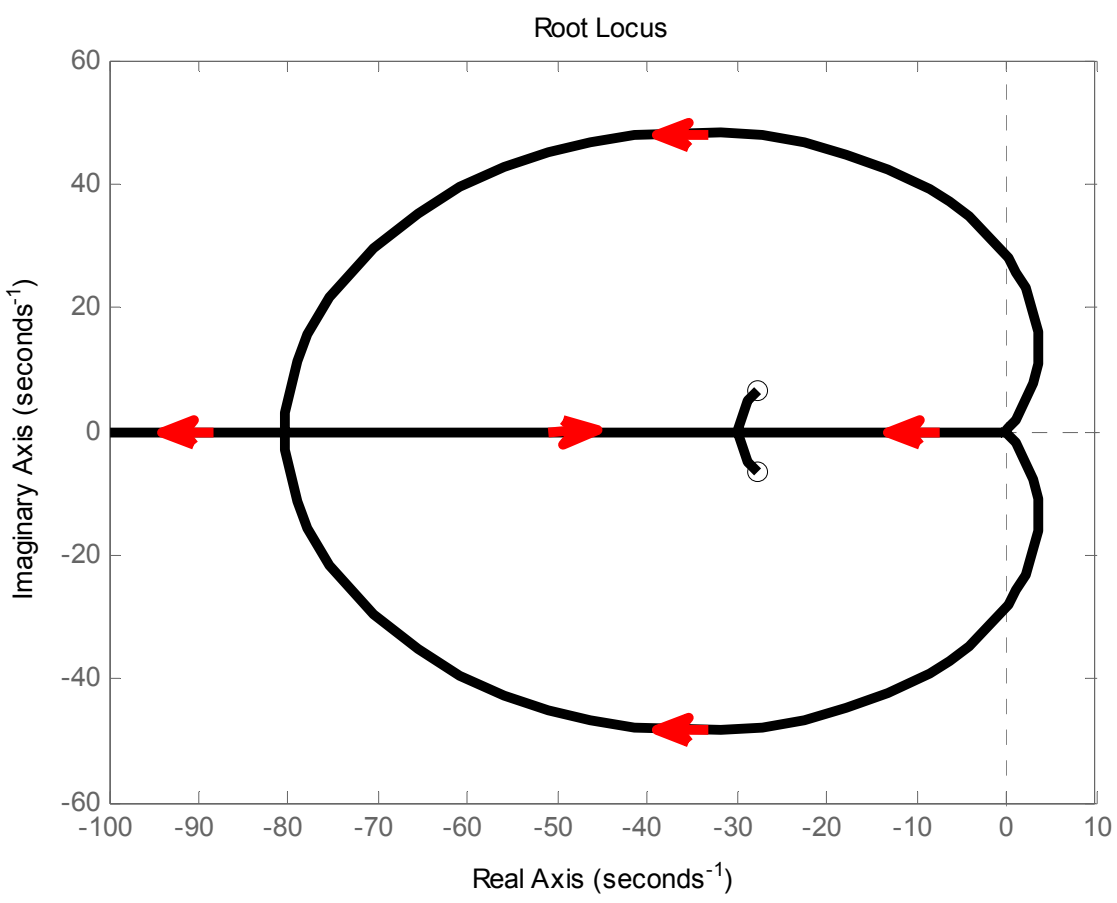

Fig. 7. Stability of the robust position control system

7. It is clear from the figure that the stability of the robust position control system improves as $\alpha$ is increased. In the Fig. 8, the root-loci of the robust force control system are plotted with respect to the force control gain $C_f$. The Fig.8 a shows that the stability of the robust force control system deteriorates as $C_f$ is increased; increasing the bandwidth of a RFOB improves the stability. The Fig 8b shows that the stability of the robust force control system deteriorates due to right half plane zero if $\hat{J}_m > J_m$. To improve the stability, $\hat{J}_m \le J_m$ should be guaranteed in the design of a RFOB.

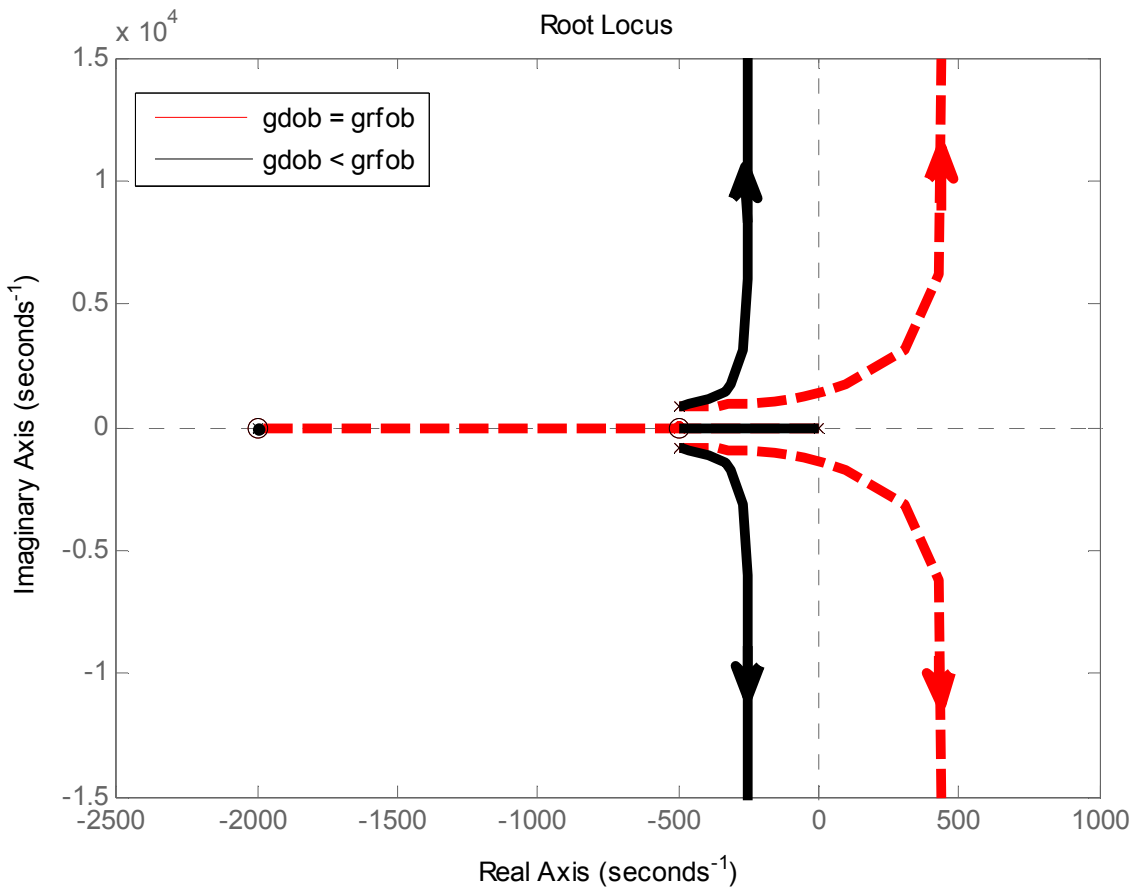

(a) $\hat{J}_m = J_m$

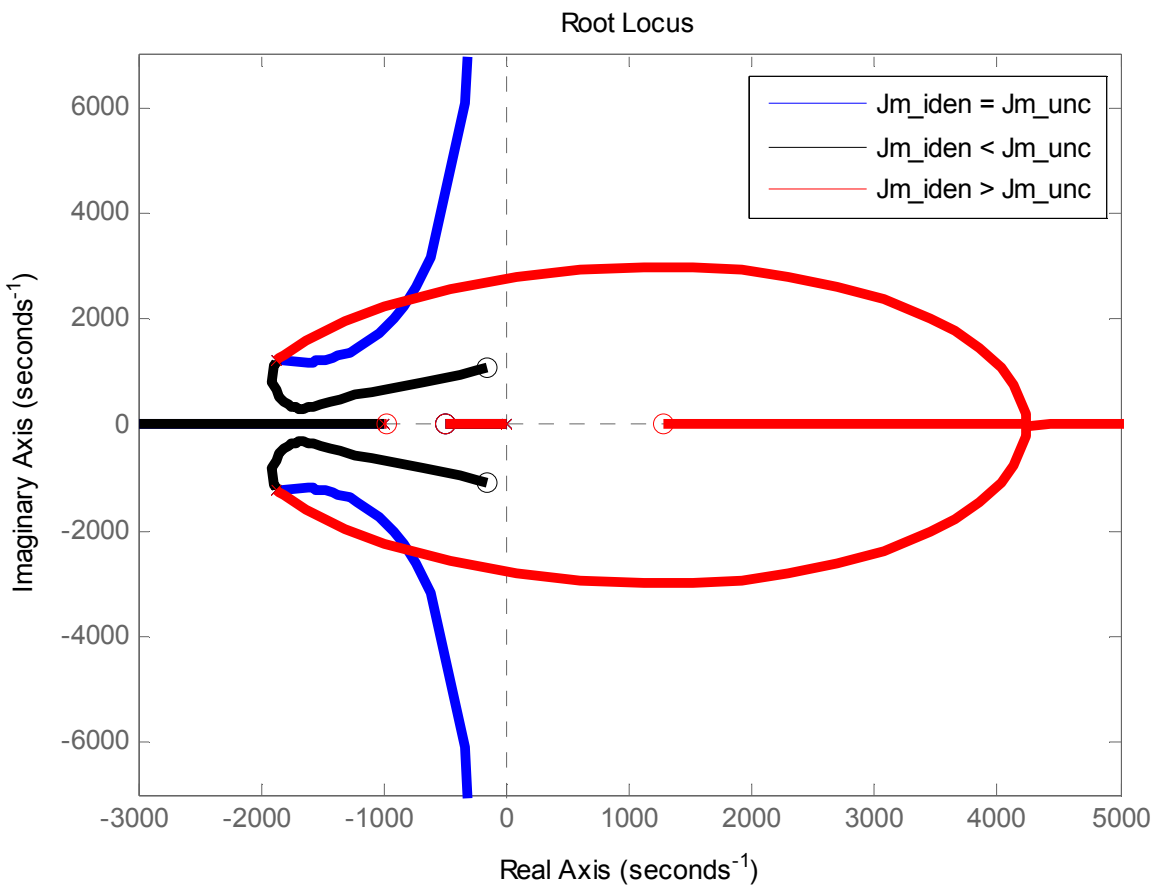

(b) $g_{RFOB} > g_{DOB}$

Fig. 8 Stability of the robust force control system

Fig. 9 shows the position control responses of the DC motor when a DOB is implemented. In this experiment, a sinusoidal position reference input is applied between 1 to 10 seconds, and the position control responses are observed by changing the nominal inertia in the design of the DOB. The Fig 9 shows that the stability of the position control system is improved by increasing the nominal inertia, yet the noise is increased. Consequently, the trade-off between the stability and robustness is shown clearly in the Fig. 9.

Fig. 10 shows the force control responses of the DC motor when a hard environment (aluminum box) is used in the contact motion. In this experiment, the hard environment is at 0.01 m. initially, and a step force control reference is applied at 1 second. The Fig. 10 clearly shows that the stability of the robust force control system changes significantly by the design parameters of DOB and RFOB, and the stability is improved by designing $J_m > \hat{J}_m$ and $g_{RFOB} > g_{DOB}$.

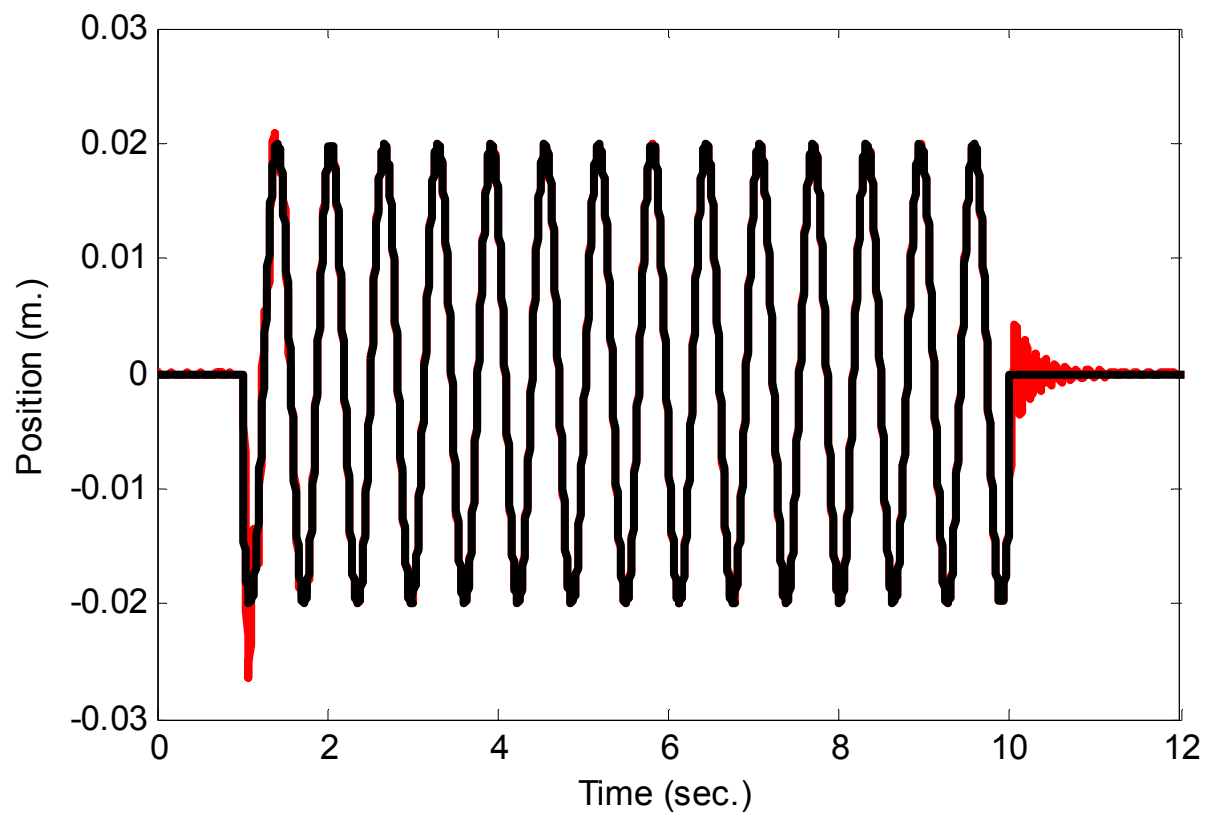

(a) Small nominal mass

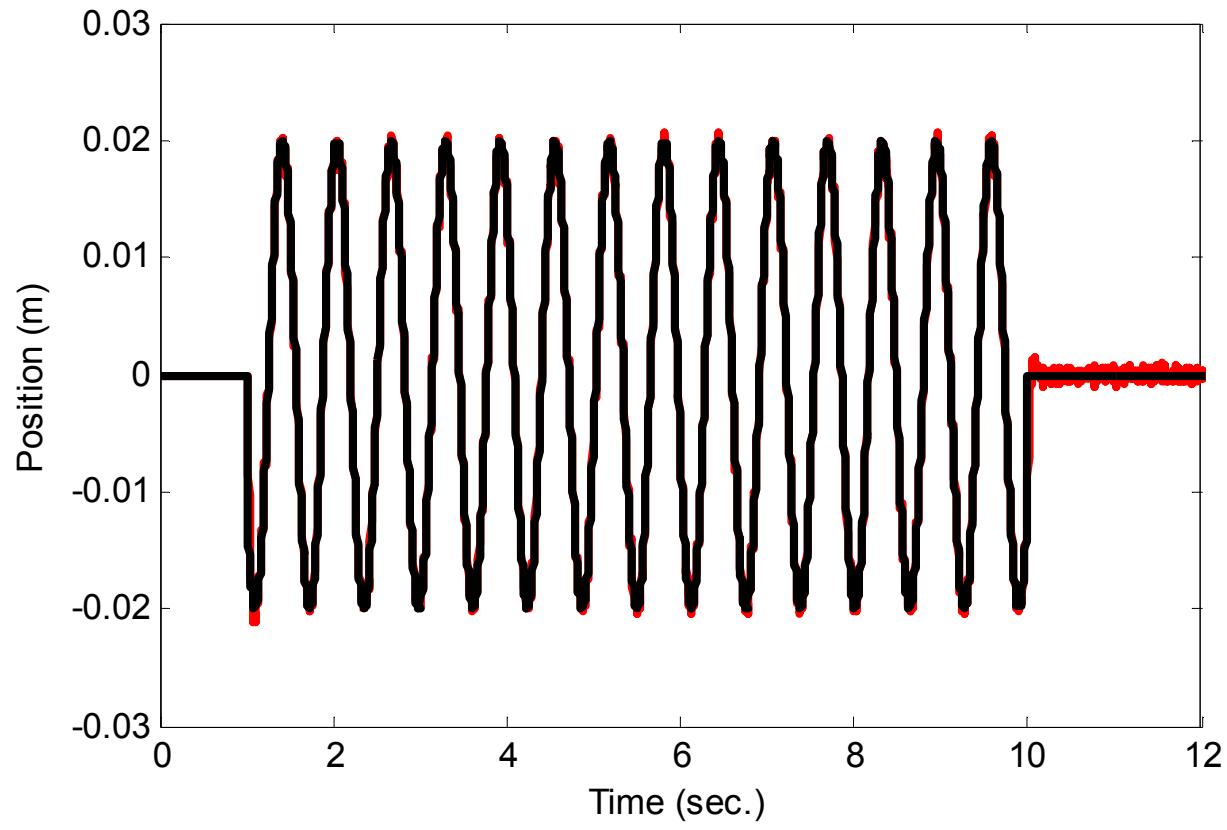

(b) Big nominal mass

Fig. 9. Position control responses

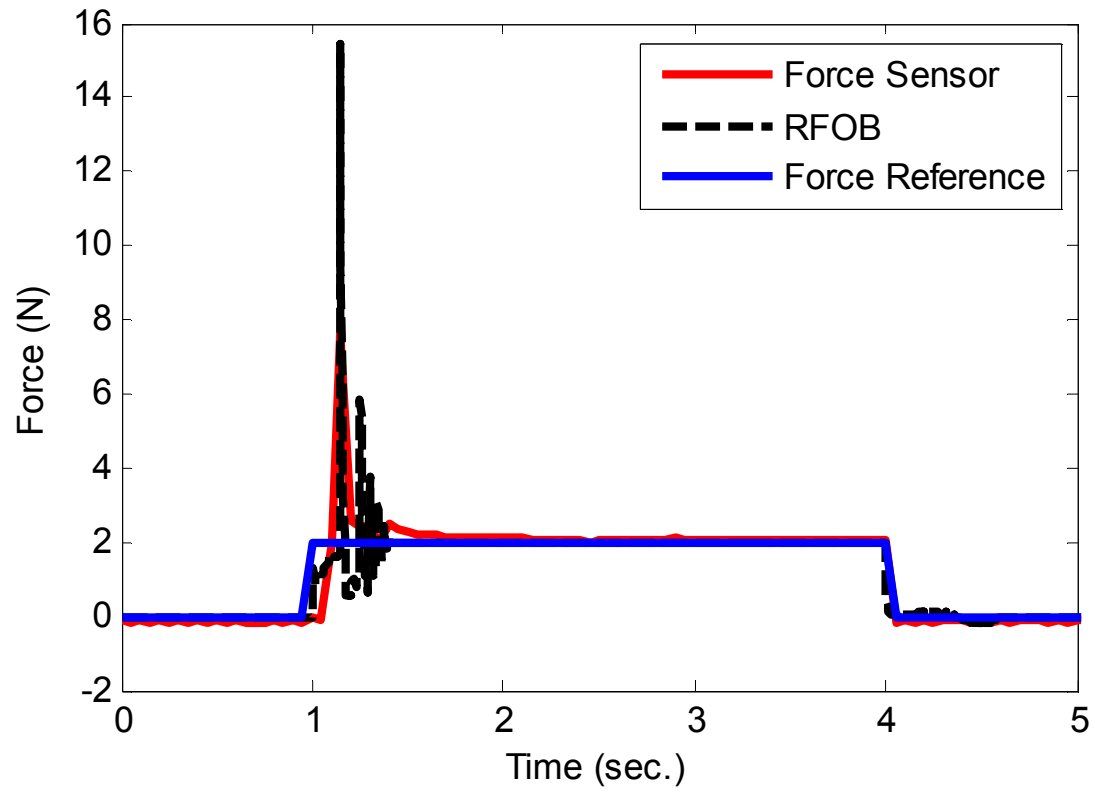

(a) $J_m \cong \hat{J}_m$ and $g_{RFOB} = g_{DOB}$

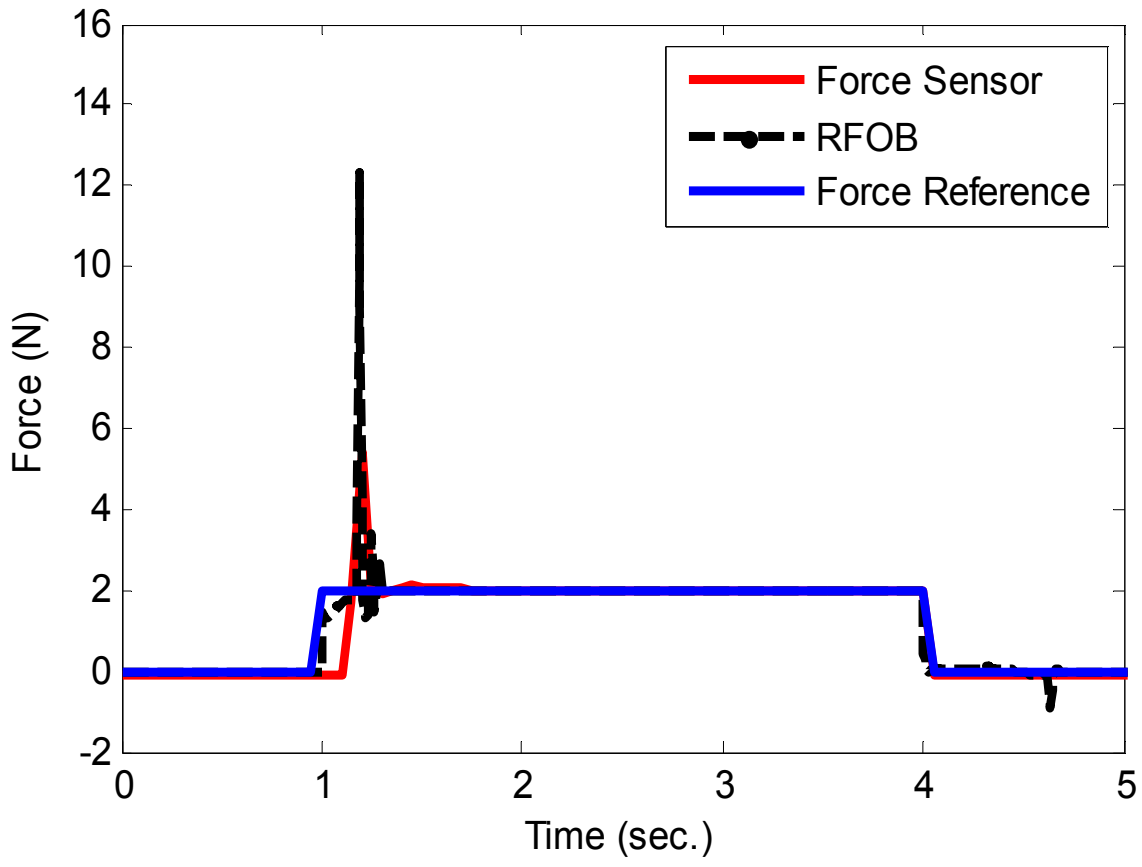

(b) $J_m > \hat{J}_m$ and $g_{RFOB} > g_{DOB}$

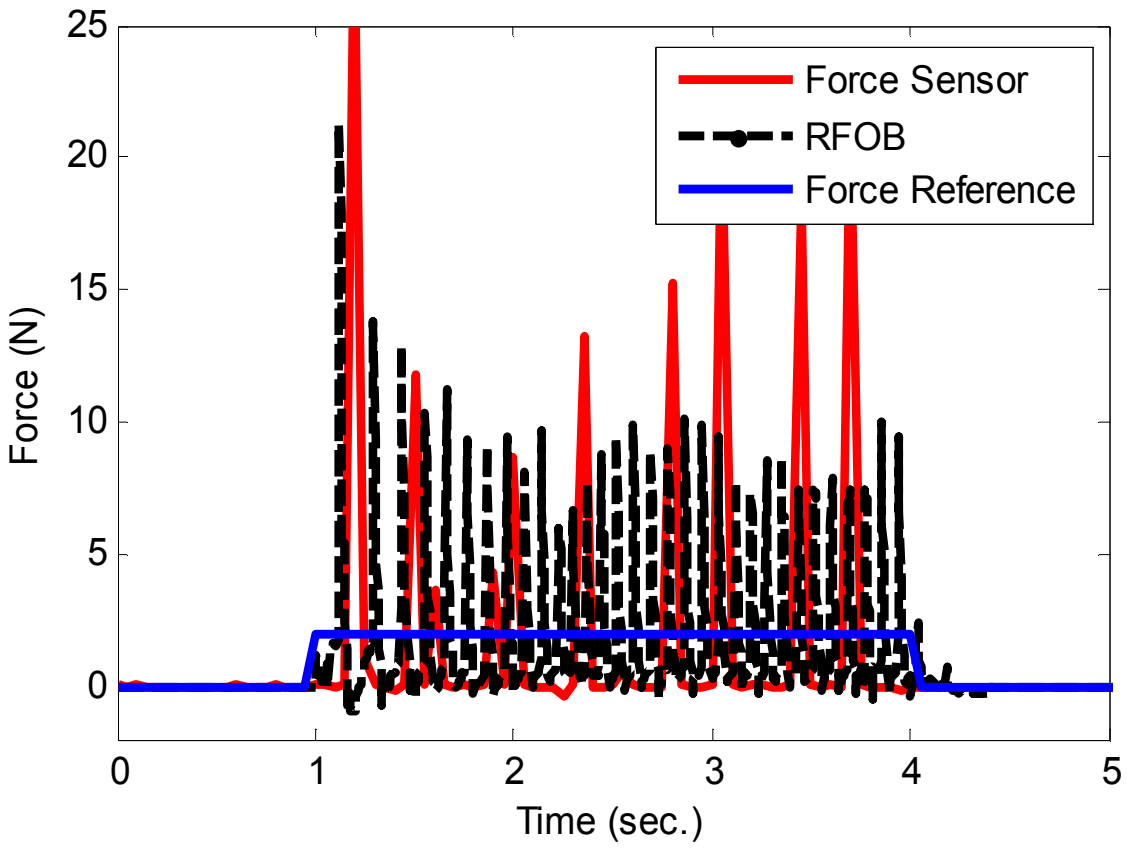

(c) $J_m < \hat{J}_m$ and $g_{RFOB} > g_{DOB}$

Fig. 10. Force control responses

## CONCLUSION

This paper proposes new design tools for the DOB and RFOB based robust position and force control systems, respectively. The proposed design tools improve the stability, robustness and performance of the robust motion control systems. By using the proposed design tools, a DOB based robust motion control system can be designed easily without requiring any pre-experiences on DOB.

The paper clarifies that velocity measurement has significant effect in the stability, robustness and performance of a DOB; and not only the performance but also the stability changes significantly by the design parameters of a DOB and a RFOB in the robust force control systems.

To improve the stability, robustness and performance of a DOB based motion control system, although the exact model of motor inertia is not required, it should be identified and a DOB and a RFOB should be designed by using the proposed constraints. Although increasing the identified torque coefficient $\hat{K}_\tau$ improves the stability of the robust force control system, the performance of the force control deteriorates significantly. Therefore, the torque coefficient should be identified precisely.

## ACKNOWLEDGEMENT

This research was supported in part by the Ministry of Education, Culture, Sports, Science and Technology of Japan under Grant-in-Aid for Scientific Research (S), 25220903, 2013.